\documentclass[journal]{IEEEtran}

\usepackage{palatino,epsfig,latexsym,,cite,graphicx,amssymb,multirow,booktabs,color,algorithm,algorithmic}
\usepackage[caption=false,font=footnotesize]{subfig}
\usepackage{threeparttable}

\usepackage{amssymb}
\usepackage{bm}

\begin{document}

\title{Detection of core-periphery structure in networks by 3-tuple motifs} 

\author{Chuang Ma, Bing-Bing Xiang, Hai-Feng Zhang, Han-Shuang Chen, and Michael Small
\thanks{This work is supported by National Natural Science Foundation of China (61473001), and partially supported by the Young Talent Funding of Anhui Provincial Universities (gxyqZD2017003). M. Small is supported by an Australian
Research Council Future Fellowship (FT110100896). \it{(Corresponding author: Hai-Feng Zhang.)}  }
\thanks{C. Ma and B.-B. Xiang are with the School of Mathematical Science, Anhui University, Hefei 230601, China (chuang\_m@126.com; xiangbb2311@163.com).}
\thanks{H.-F. Zhang is with the School of Mathematical Science, Anhui University, Hefei 230601 \& Center of Information Support  and Assurance Technology, Anhui University, Hefei, 230601, China (haifengzhang1978@gmail.com).}
\thanks{H. Chen is with the School of Physics and Material Science, Anhui University, Hefei 230601, China (chenhshf@mail.ustc.edu.cn).}

\thanks{M. Small is with the Department of Mathematics and Statistics, The
University of Western Australia, 35 Stirling Highway, Crawley, WA
6009 \& Mineral Resources, CSIRO, Kensington, Western Australia, 6151, Australia (michael.small@uwa.edu.au).}}

%



\maketitle

\begin{abstract}
Recently, the core-periphery (CP) structure of networks as one type of meso-scale structure has received attention. The CP structure is composed of a dense core and a sparse connected periphery. In this paper, we propose an algorithm to detect the CP structure based on the 3-tuple motif, which is inspired by the idea of motif. In this algorithm, we first define a 3-tuple motif by considering the property of nodes, and then a motif adjacency matrix is formed based on the defined motif, finally, the detection of the CP structure is converted to find a cluster that minimizes the smallest motif conductance. Our algorithm can detect different CP structures: including single or multiple CP structure; and local or global CP structures. Results in the synthetic and the empirical networks indicate that the method is efficient and can apply to large-scale networks. More importantly, our algorithm is parameter free, where the core and periphery are detected without the need for any predefined parameters.
\end{abstract}


\section{Introduction} \label{sec1}

The structure of networks can be described from micro-scale, meso-scale and macro-scale perspectives. Recently, myriad methods have been developed to detect one typical meso-scale structure: community structure~\cite{lu2015algorithms,fortunato2010community,mahmood2016subspace,li2016fast}. However, another type of meso-scale structures: the core-periphery (CP) structure, has received much less attention.
The CP structure has been examined in the study of society~\cite{white1976social}, transportation~\cite{verma2016emergence}, scientific citation~\cite{doreian1985structural}, international trade~\cite{nemeth1985international,smith1992structure} and other fields~\cite{doolittle1996self,barucca2016disentangling,holme2005core}.  Although a core and a community are both a group
of densely interconnected nodes,  there is a remarkable difference between them,  since a core should also be densely connected the periphery~\cite{kojaku2017finding}.

Though the intuitive notation of CP has been mentioned for some time~\cite{snyder1979structural}, the first quantitative formulation of CP structure was proposed by Borgatti and Everett in the late 1990s ~\cite{borgatti2000models}, in which a node belongs to a core if and only if it is well connected both with other core nodes and with peripheral nodes, and peripheral nodes have no connections with other peripheral nodes. Hereafter, some detection algorithms of CP structure have been proposed. There are several ways of design CP detection algorithms: one is to maximize the objective functions characterizing the similarity between the partition and its corresponding idealised CP structure~\cite{borgatti2000models,rombach2014core,kojaku2017finding}. Another is that the CP structure is detected according to a certain ordering sequence encoding the core value of each node~\cite{da2008centrality,shanahan2012knotty,della2013profiling,cucuringu2014detection,gamble2016node}. In addition, Zhang \emph{et~al.} have proposed that the CP structure can be detected by fitting a stochastic block model (SBM) to empirical
network data using a maximum likelihood method~\cite{zhang2015identification}. Many proposed algorithms can only deal with a single CP structure, require that the size of core nodes should be given in advance, or have high computational complexity.

Recently, Benson \emph{et~al.} developed a generalized framework for clustering
networks on the basis of structural motifs instead of edges~\cite{benson2016higher}. The motif is described by a 2-tuple $(B, \cal A)$, where $B$ is a $k\times k$ binary matrix and ${\cal A} \subset \left\{ {1,2, \ldots ,k} \right\}$ is a set of anchor nodes. They then proved that the near optimal clustering of the network can be realized by finding the set with the smallest motif conductance. Such a method can be applied to directed, undirected, weighted networks, signed networks, and so forth. However, the motifs defined in this work primarily focused on the patterns of edges in the subgraph, which has no restriction of the nodes of the subgraph. So we ask: is it possible to uncover higher-organization structures of networks by define some motifs which incorporate the information of nodes.

From the definition of CP structure, it is known that core nodes are the high-degree nodes, even though the high-degree nodes are not necessarily the core nodes~\cite{borgatti2000models,zhang2015identification}. Inspired by this fact, we first defined motifs by a 3-tuple $(B, {\cal A} , \Phi)$. The property $\Phi$ is to highlight the degrees of core nodes are larger than the peripheral nodes, and also larger than the average degree of the network. As in Ref.~\cite{benson2016higher}, the motif adjacency matrix is first constructed on the basis of the defined motif, then the dichotomization of the network is implemented by finding a cluster that minimizes the smallest motif conductance. Finally, the subgraph with larger average degree is the core, and the other is the periphery. We further generalize our algorithm to detect multiple cores and the global core, respectively. The performance of the algorithm is validated in different networks. Importantly, the algorithm is fast and parameter free.

%
%
\section{Related work} \label{sec_new}

The detection of CP structure was formally considered by Borgatti and Everett~\cite{borgatti2000models}. In that work, a discrete algorithm aims to find a vector $C$ of length $N$ whose entries can be either 1 or 0. The $i$th entry $C_i$ equals 1 if the
corresponding node is assigned to the core, otherwise, it equals 0 if the node is
assigned to the periphery. Next define a pattern matrix with the element $C_{ij}=1$ if $C_i=1$ or $C_j=1$, and let $C_{ij}=0$ otherwise. A core quality $\rho=\sum_{i,j}A_{ij}C_{ij}$ is defined to measure how well a network approximates the ideal case, where $A_{ij}$ is an element of the adjacency matrix $A$. The goal of the algorithm is to seek a value of  $\rho$ that is high compared to the
expected value of $\rho$ if $C$ is shuffled such that the number of 1 and 0 entries are
preserved but their order is randomized~\cite{borgatti2000models,rombach2014core,cucuringu2014detection}. They also presented a continuous version in which each node is assigned a ``coreness'' value between 0 and 1, and the element of the pattern matrix $C_{ij}=C_iC_j$. The continuous algorithm was further developed in Ref.~\cite{rombach2014core}, where the aggregate core score of each node is defined by the core quality as well as the transition function. Very recently, the problem of CP detection is further developed by Kajoku \emph{et~al.}~\cite{kojaku2017finding}, they define a more general objective function to find multiple CP pairs in networks, the multiple CP pairs can be found by maximizing the quality function.

Some methods detect core nodes by ranking nodes according to a defined core centrality. For instance, based on the fact that core nodes should have a high closeness centrality, Holme~\cite{holme2005core} proposed a CP coefficient using the closeness centrality and k-cores deposition technique to determine core nodes. Silva \emph{et~al.}~\cite{da2008centrality} proposed a parameter called core coefficient to quantitatively evaluate the core-periphery structure of a network, which was defined based on the concept of closeness centrality and a newly defined parameter: network capacity. A knotty centrality was proposed by Shanahan \emph{et~al.}~\cite{shanahan2012knotty} to detect the core in networks, which attempts to find nodes that have high betweenness centrality but without high degree centrality. In addition, other methods to measure the core values of nodes, such as, based on the path-core~\cite{cucuringu2014detection} or random walk~\cite{della2013profiling}, were studied.

Methods to maximize the quality/objective function characterizing the similarity between a given partition and its corresponding idealised CP structure may yield inefficient results, since empirical networks often significantly deviate from the idealised CP structure. Given that, Zhang \emph{et al.}~\cite{zhang2015identification} proposed an algorithm to identify the CP structure by fitting a stochastic block model to empirical network data using a maximum likelihood method, rather than the idealised CP structure.

Our algorithm is totally different from these previous approaches. In our algorithm, we first generalize the definition of 2-tuple motif to 3-tuple motif by considering the property of nodes, and then a motif adjacency matrix is constructed based on the 3-tuple motif, finally, the detection of the CP structure is converted to find a cluster that minimizes the smallest motif conductance. Our algorithm can detect single or multiple CP structure, as well as local or global CP structures.

\section{Motif Spectral Clustering} \label{sec2}

We first review the method developed in Ref.~\cite{benson2016higher}, which is to detect the higher-order organizations by finding a cluster with the smallest motif conductance. There are three main steps to detect higher-order organizations:\\

\begin{figure}
\begin{center}
\includegraphics[width=3.2in]{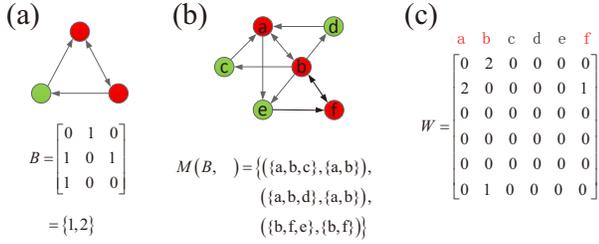}
\caption{Diagram of motif definition $M$ and construction of motif adjacency matrix $W_{M}$. (a) Diagram of motif definitions. The motif is defined by a binary matrix $B$ and an anchor set of nodes, and the red anchors are connected by a bi-directional edge. (b) Given the  motif definition $M$ in (a), three instances of the motif can be found in the center graph. (c) The elements of the motif adjacency matrix ($W_M$) are determined by counting the number of times two nodes co-occur in an instance of the motif.}
\label{fig1}
\end{center}
\end{figure}

\textbf{Define a motif $M$ and form the motif adjacency matrix $W_{M}$:} Consider an undirected graph $G = (V,E)$ with $|V| = n$, the definition of motif is described by a 2-tuple $\left( {B,{\cal A}} \right)$ on $k$ nodes. The binary matrix $B$ includes the patterns of edges in the subgraph. ${\cal A} \subset \left\{ {1,2, \ldots ,k} \right\}$ is a set of anchor nodes, which labels a relevant subset of nodes for defining motif conductance (see Fig.~\ref{fig1}(a)).  Let ${\chi _{\cal A}}$ be a selection function that takes the subset of a $k$-tuple indexed by ${\cal A}$, and $set\left(  \bullet  \right)$ be the operator taking a tuple to a set. Namely,
\begin{eqnarray}\label{1}
set\left( {\left( {{v_1},{v_2}, \cdots ,{v_k}} \right)} \right) = \left\{ {{v_1},{v_2}, \cdots ,{v_k}} \right\}.
\end{eqnarray}
Assume a network $G$ is denoted by an adjacency matrix $A$, then the set of motifs is defined as:
\begin{eqnarray}\label{2}
\nonumber M( {B,{\cal A}}) &=& \{(set(\textbf{v}), set(\chi_{{\cal A}}(\textbf{v}))|
\\&& \textbf{v}\in V^k,~v_1,\cdots,v_k ~distinct, A_\textbf{v}=B \}.
\end{eqnarray}
where ${{A_{\bf{v}}}}$ is the $k \times k$ adjacency matrix on the subgraph induced by the $k$ nodes of the ordered vector ${\bf{v}}$. For convenience, $\left( {set\left( {\bf{v}} \right), set\left( {{\chi _{\cal A}}\left( {\bf{v}} \right)} \right)} \right)$ is simply denoted as $\left( {{\bf{v}},{\chi _{\cal A}}\left( {\bf{v}} \right)} \right)$ when  considering the elements of $M\left( {B,{\cal A}} \right)$. Moreover, any $\left( {{\bf{v}},{\chi _{\cal A}}\left( {\bf{v}} \right)} \right) \in M\left( {B,{\cal A}} \right)$ is called a motif instance (see Fig.~\ref{fig1}(b)).

Once the adjacency matrix $A$ and a motif set $M$ are determined, the motif adjacency matrix (Fig.~\ref{fig1}(c)) can be formally defined as:
\begin{eqnarray}\label{3}
{\left( {{W_M}} \right)_{ij}} = \sum\limits_{\left( {{\bf{v}},{\chi _A}\left( {\bf{v}} \right)} \right) \in M} {\textbf{1}\left( {\left\{ {i,j} \right\} \in {\chi _A}\left( {\bf{v}} \right)} \right)},
\end{eqnarray}
where $\textbf{1}(s)$ is the truth-value indicator function on $s$, i.e., $\mathbf{1}(s)$ takes the value 1 if the statement
$s$ is true and 0 otherwise.

\textbf{Apply spectral clustering on $W_M$:}
Given a motif $M$, an optimal cluster of nodes $S$ has two goals: on the one hand, the nodes in $S$ should contain many instances of $M$; on the other hand, the set $S$ should avoid cutting instances of $M$. Thus, the optimal cluster is realized by minimizing the following motif conductance:
 \begin{eqnarray}\label{4}
{\phi _M}\left( S \right) = \frac{{cu{t_M}(S,\bar S)}}{{\min \left( {vo{l_M}\left( G \right)S,vo{l_M}\left( G \right)\bar S} \right)}},
 \end{eqnarray}
where ${\bar S}$ is the complement of $S$, namely, $\bar{S}=V\setminus S$. Let ${cut{_M}(S,\bar S)}$ denote the number of instances of motif $M$ with at least one node in $S$ and one in ${\bar S}$, and ${vo{l_M}\left( G \right)S}$ the number of nodes in instances of $M$ that reside in $S$. The supplementary materials of Ref.~\cite{benson2016higher} proves that the near optimal clusters can by obtained by using motif spectral clustering on the motif adjacency matrix $W_M$.

First, define the normalized Laplacian matrix as:
 \begin{eqnarray}\label{5}
L_M = I - D_M^{ - 1/2}{W_M}D_M^{ - 1/2},
 \end{eqnarray}
where ${\left( {{D_M}} \right)_{ii}} = \sum\limits_{j = 1}^n {{{\left( {{W_M}} \right)}_{ij}}}$ is the diagonal degree matrix and $I$ is the identity matrix. Then compute the eigenvector of the second smallest eigenvalue of $L_M$, denoted by $z$. Let $\sigma_i$ be the index of $D_M^{ - 1/2}z$ with the $i$th smallest value.

\textbf{Output the clusters:} After the vector $\sigma=\{\sigma_1,\sigma_2,\cdots,\sigma_n\}$ is given, the final step is to find the prefix set of $\sigma$ with the smallest motif conductance, namely, $S=\arg \min \phi _M( S_r)$ with $S_r=\{\sigma_1,\sigma_2, \cdots ,\sigma_r\}$.
The detail steps are presented in Algorithm 1.\\

\textbf{Algorithm 1:} (Input: graph $G$ and motif $M$. Output: optimal cluster $S$)

\textbf{Step 1:} Compute $W_M$ according to Eq.~(\ref{3});

\textbf{Step 2:} Compute the normalized motif Laplacian  matrix according to Eq.~(\ref{5});

\textbf{Step 3:} Compute the eigenvector $z$, which is the second smallest eigenvalue of ${L_M}$;

\textbf{Step 4:} Let ${\sigma _i}$ be the index of $D_M^{ - 1/2}z$ with the $i$th smallest value;

\textbf{Step 5:} Find $S = \arg \min {\phi _M}\left( {{S_r}} \right)$ by increasing size $r$, where $S_r=\{\sigma_1,\sigma_2, \cdots ,\sigma_r\}$;

\textbf{Step 6:} If $\left| S \right| < \left| {\bar S} \right|$, then return $S$, else return ${\bar S}$.

From the description of Algorithm 1, one can find that there is an assumption in Step 6: the optimal clustering $S$ is the smallest set of $S$ and ${\bar S}$. However, there is no evidence to support this assumption. Moreover, according to this algorithm, one optimal cluster $S$ can be detected, however, the cluster will be wrong if in fact there is no optimal cluster. This problem is very similar to the detection of communities in networks. The modularity optimization algorithm can produce a wrong partition even though one network has no obvious community structure. As a result, some improvements can be made in this method. In this work, we generalize the motif spectral clustering method to detect the CP structure in networks, which also overcomes these mentioned shortcomings.

\section{Methods} \label{sec3}
The definition of motif in the above section merely considers the patterns of edges, but ignores the property of the nodes. In many cases, the property of the nodes is also important, and considering the property of nodes may help us find more important organizations. For example, given that the core nodes in the network must be the high-degree nodes, therefore, we define a new type of motif by considering the degree property of nodes. Namely, motif is extended into a 3-tuple: $\left( {B,{\cal A},\Phi} \right)$, where $\Phi$ is the property of nodes.

By extending a 2-tuple motif definition to a 3-tuple motif definition, which is defined as:
\begin{eqnarray}\label{6}
\nonumber M( {B,{\cal A}},\Phi) &=& \{(set(\textbf{v}), set(\chi_{{\cal A}}(\textbf{v}))|
\\&& \nonumber\textbf{v}\in V^k,~v_1,\cdots,v_k ~distinct, A_\textbf{v}=B, \\&& f(\Phi,(set(\bf{v}), set(\bf{\chi_{\cal A}}(v))))=1 \}.
\end{eqnarray}
Where $f\left( {\Phi,\left( {set\left( {\bf{v}} \right),set\left( {{\chi _{\cal A}}\left( {\bf{v}} \right)} \right)} \right)} \right) = 1$  indicates that the node of $set\left( {\bf{v}} \right)$ and $set\left( {{\chi _{\cal A}}\left( {\bf{v}} \right)} \right)$ should satisfy the property of $\Phi$.

\begin{figure}
\begin{center}
\includegraphics[width=3.2in]{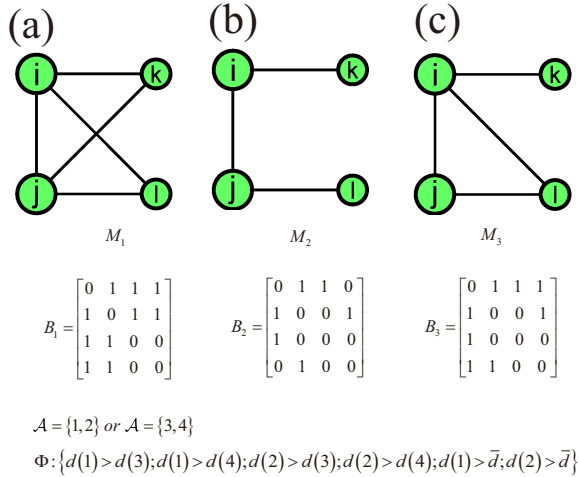}
\caption{Diagram of 3-tuple motif definitions. (a) motif ${M_1} = \left( {{B_1},{\cal A},\Phi} \right)$. (b) motif ${M_2} = \left( {{B_2},{\cal A},\Phi} \right)$. (c) motif ${M_3} = \left( {{B_3},{\cal A},\Phi} \right)$. Where $d( \bullet )$ is the degree of a node, $\bar d$ is the average degree, $M$ is a core motif when ${\cal A} = \left\{ {1,2} \right\}$ and is a periphery motif when ${\cal A} = \left\{ {3,4} \right\}$. $\Phi$ is to guarantee the degrees of core nodes are larger than the peripheral nodes, and further larger than the average degree of the networks.}
\label{fig2}
\end{center}
\end{figure}

According to the definition of CP in Ref.~\cite{borgatti2000models}, one core node should connect with all other nodes, and a peripheral node should connect with all core nodes but does not connect with other peripheral nodes. To meet this condition, a 3-tuple motif is defined in Fig.~\ref{fig2}(a), which is determined by the binary matrix $B_1$ (or $B_2$, $B_3$), the set of anchor nodes is ${\cal A}=\{1, 2\}$ or $\{3, 4\}$, and $\Phi: \{d(1)>d(3);d(1)>d(4);d(2)>d(3);d(2)>d(4);d(1)>\bar{d};d(2)>\bar{d}\}$, here $\bar{d}$ is the average degree of the network.

\begin{figure}
\begin{center}
\includegraphics[width=3.2in]{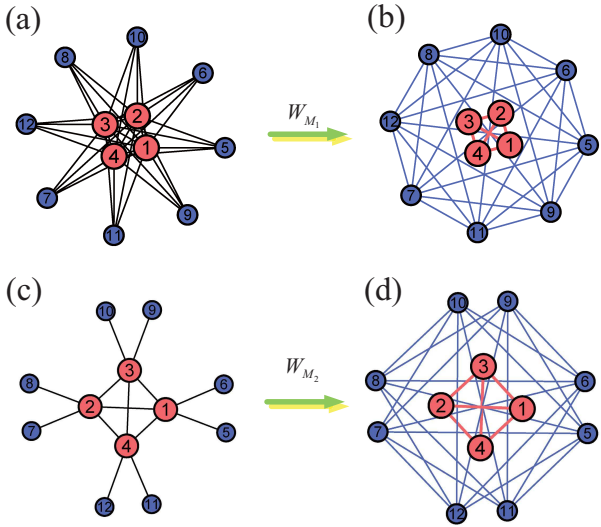}
\caption{The topological transformations  based on different motif adjacency matrices. Core nodes and peripheral nodes are colored by red and blue, respectively. The size of nodes represent degrees and the thickness of edges represent weights. (a) and (b) show that matrix $W_{M_1}$ can dichotomize a classical CP structure into  two separated well-connected subgraphs: core and periphery;  (c) and (d) show that matrix $W_{M_2}$ can dichotomize a looser defined CP structure into two separated well-connected subgraphs: core and periphery.}
\label{fig3}
\end{center}
\end{figure}

Let $M_1^C = \left( {{B_1},{{\cal A}^C},\Phi} \right)$ and $M_1^P = \left( {{B_1},{{\cal A}^P},\Phi} \right)$ be the motif of core and motif of periphery, respectively, where ${{\cal A}^C} = \{ 1,2\} $ and ${{\cal A}^P} = \{ 3,4\} $. Then the motif adjacency matrix of $M_1$ is formed as:
\begin{eqnarray}\label{7}
 {\left( {{W_{{M_1}}}} \right)_{ij}}&& = \sum\limits_{\left( {{\bf{v}},{\chi _{\cal A}}\left( {\bf{v}} \right)} \right) \in M_{_1}^C\bigcup M_{1}^P} {\textbf{1}\left( {\left\{ {i,j} \right\} \in {\chi _{\cal A}}\left( {\bf{v}} \right)} \right)} .
\end{eqnarray}
Note that the definition of $W_{M_1}$ in Eq.~(\ref{7}) is different from Eq.~(\ref{3}), which is defined to ensure the weight among core nodes and the weight among peripheral nodes are increased, however, the weight between core nodes and peripheral nodes are not changed. Therefore the difference between core and periphery is more obvious (see the diagrams in Fig.~\ref{fig3}). Namely, such a definition is in favor of the detection of CP structure. For example, take the classical CP structure in Fig.~\ref{fig3}(a) as an example, the network is divided into two well-connected subgraphs (see Fig.~\ref{fig3}(b)) after the function mapping of $W_{M_1}$: core and periphery. From this example, one can also find that our method need not use the assumption in step 6 of Algorithm 1 (the optimal cluster is the smallest set of $S$ and $\bar{S}$). We only need to judge whether the degrees of all nodes in $S$ or $\bar{S}$ are larger than the average degree $\bar{d}$. Obviously, a node which satisfies this condition is the core, and the other is the periphery.

In many cases, the CP structure criterion shown in Fig.~\ref{fig3}(a) is too strict to  be met. The diagram in Fig.~\ref{fig3}(c) can also be viewed as a CP structure. If we still construct the motif adjacency matrix based on $M_1$, the CP structure in Fig.~\ref{fig3}(c) cannot be detected since $W_{M_1}$ is a zero-matrix. To do this, we define a motif $M_2$ as in Fig.~\ref{fig2}(b), then the motif adjacency matrix $W_{M_2}$ is formed according to Eq.~(\ref{7}). As shown in Fig.~\ref{fig3}(d),  the illustration in Fig.~\ref{fig3}(c) can be divided two separated sets after the function mapping of ${W_{{M_2}}}$: core and periphery.

Since the definition of motif $M_1$ is too strict and the definition of motif $M_2$ is very loose, a modest motif $M_3$ intermediated motif $M_1$ and $M_2$ is also defined (see Fig.~\ref{fig2}(c)). It is hard to know whether the CP structure is obvious in a given network owing to the richness and diversity of real networks. In view of this, we define motif adjacency matrix $M_W$ as the combination of $W_{M_1}$, $W_{M_2}$ and $W_{M_3}$:
\begin{eqnarray}\label{8}
{W_M} = \alpha {W_{{M_1}}} + \gamma {W_{{M_2}}} + \beta{W_{{M_3}}},
\end{eqnarray}
where $\alpha  \ge \beta  \ge \gamma  \ge 0$. Since motif $M_2$ is encoded in motif $M_3$, and further encoded in motif $M_1$. In this paper, without loss of generality, we set $\alpha  = 2\beta$,  $\beta  = 2\gamma $ and $\gamma =1.0$.

In summary, the motif adjacency matrix is first obtained according to Eq.~(\ref{8}) for a given network, then $S$ and $\bar S$ will be obtained by Step 1 - Step 5 in Algorithm 1. Finally, the core is the set $S$ or $\bar S$ in which degrees of all nodes are larger than the average degree. In the next section, we validate the performance of our algorithm by implementing it in several real networks and synthetic networks.
\section{Main results} \label{sec4}
\subsection{Detection of single CP structure}\label{sec4.1}

\begin{figure*}
\begin{center}
\includegraphics[width=6in]{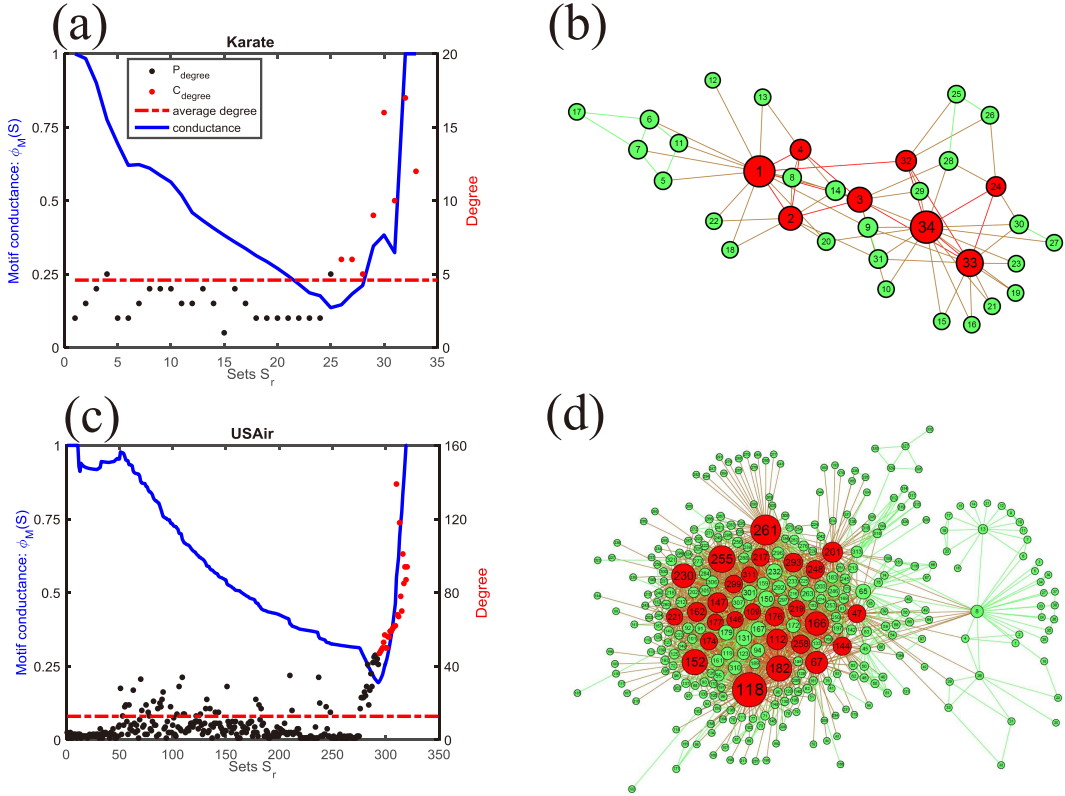}
\caption{Detection of single CP structure in the Karate club network and the USA airport network based on Algorithm 1. The motif conductance ${\phi _M}\left( {{S_r}} \right)$ as a function of ${S_r}$ for the two networks are shown in (a) and (c), respectively, which is shown in blue curve and is calculated from Eq.~(\ref{4}). ${S_r} = \left\{ {{\sigma _1},{\sigma _2}, \ldots ,{\sigma _r}} \right\}$ is computed from the Step 1-Step 4 of Algorithm 1.
The red dashed line is the average degree of the network. Red nodes and the black nodes are the core nodes and the peripheral nodes, respectively. Left and the right vertical ordinates denote the value of ${\phi _M}\left( {{S_r}}\right)$ and the node's degree, respectively.
Subfigures (b) and (d) are the visualizations of the Karate club network and the USA airport network, respectively. Red nodes and green nodes in (b) and (d) are the core and periphery, respectively. The size of each
node is proportional to its degree.}
\label{fig5}
\end{center}
\end{figure*}

 We first validate our algorithm on two real networks. The first network is the Karate club network, which consists of 34 nodes that represent club members and 78 links that represent friendships among members~\cite{girvan2002community}, whose motif conductance ${\phi _M}\left( {{S_r}} \right)$ (blue line) is shown in Fig.~\ref{fig5}(a). The minimum value of ${\phi _M}\left( {{S_r}} \right)$  dichotomizes the network: core and periphery. The eight red nodes in the right are the core nodes since their degrees are all larger than the average degree of the network (dashed line), and the remainder black nodes in the left are the peripheral nodes. Previous studies have found that the Karate club network is a community network including two communities. One possible scenario is that the network has two separate cores, where each community includes one core. However, our algorithm detects one core since there are some links connecting the core nodes (red nodes and red edges, as shown in Fig.~\ref{fig5}(b)), which can merge two small cores into a larger core, leading to a single CP structure in the network.

 The second network is the USA airport network, which has 332 nodes representing airports and 2126 unweighted links describing the scheduled flight between airports~\cite{lu2011link}. The blue curve in Fig.~\ref{fig5}(c) plots the motif conductance ${\phi _M}\left( {{S_r}} \right)$ as a function of $S_r$.  Furthermore, the smallest motif conductance ${\phi _M}\left( {{S_r}} \right)$ can automatically dichotomize the network as two sets: core and periphery. The core is composed of 27 nodes whose degrees are larger than the average degree (red nodes in Fig.~\ref{fig5}(c)), another set is the periphery (black nodes in Fig.~\ref{fig5}(c)). The illustration of the CP structure is presented in Fig.~\ref{fig5}(d), where core nodes and peripheral nodes are labeled by red color and green color, respectively.

\subsection{Detection of multiple cores} \label{sec4.2}

\begin{figure*}
\begin{center}
\includegraphics[width=6in]{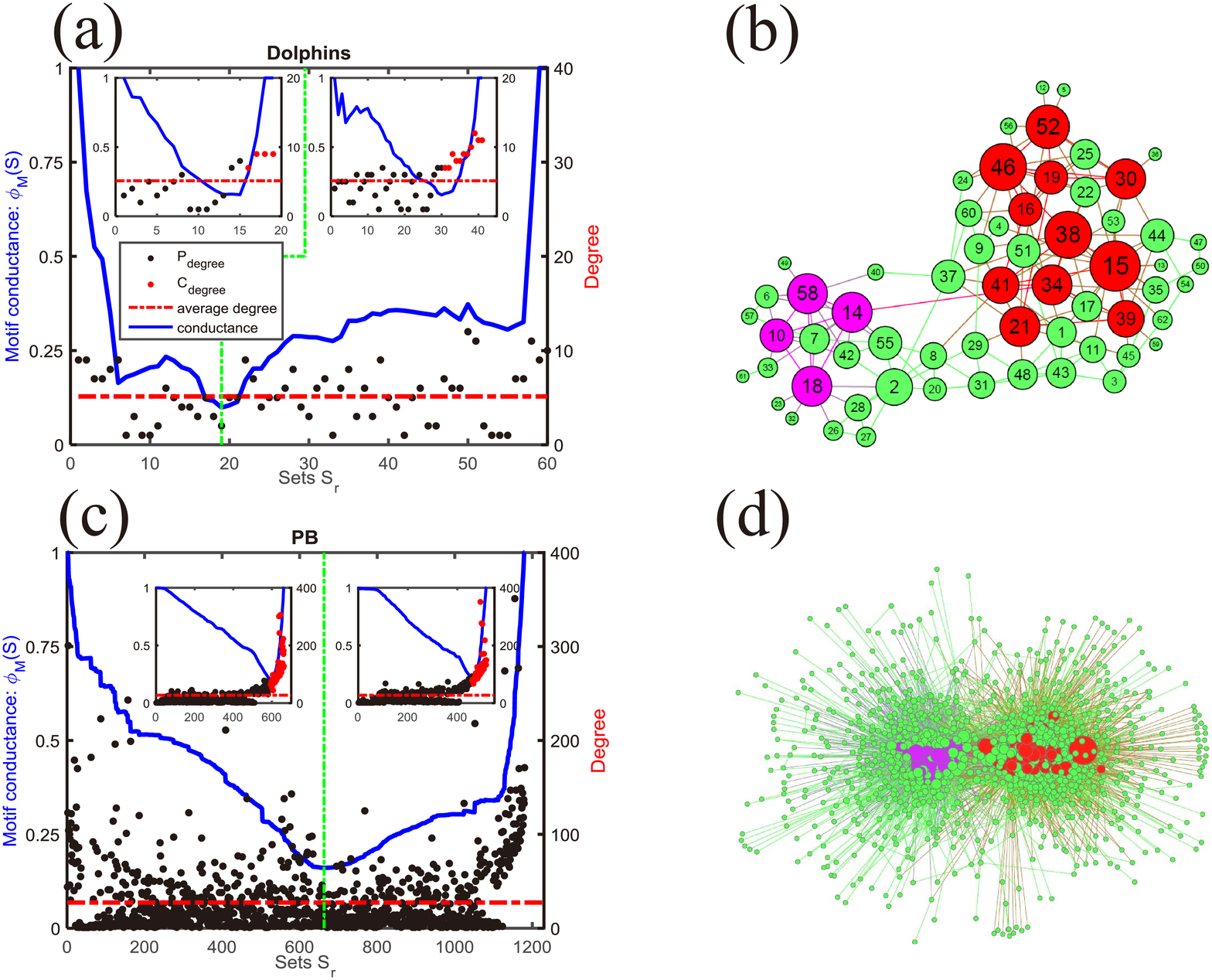}
\caption{Detection of multiple cores in Dolphin social network and Political blogs network based on Algorithm 2. (a) and (c) correspond to Dolphin social network and Political blogs network, respectively, which are plotted as in Fig.~\ref{fig5}(a) and (c). The insets of both figures are the second dichotomization on the two subgraphs: $S$ and $\bar{S}$, which are obtained by the first dichotomization (labeled by dashed green line). (b) and (d) are the visualizations of Dolphin social network and Political blogs network, respectively.  Red nodes and purple nodes are two separated cores, and the green nodes are periphery. The size of a node is proportional to its degree}
\label{fig6}
\end{center}
\end{figure*}

According to Algorithm 1, it is possible that even though the smallest motif conductance ${\phi _M}\left( {{S_r}} \right)$ can dichotomize the network into two sets, neither of them satisfies the condition that the degrees of all nodes are larger than the average degree (see Fig.~\ref{fig6}(a) and (c)). This scenario may occur for two reasons: the network itself does not exhibit CP structure at all, or the detection of the CP structure is influenced by the community structure. Because core nodes in different communities are sparsely connected \cite{girvan2002community,newman2004finding}, the two sets obtained by our Algorithm 1 may be two communities rather than the core and periphery. In this case, the CP structure may be encoded in the communities. Therefore, our algorithm is further developed in the following.

If a network includes community structure, the smallest motif conductance ${\phi _M}\left( {{S_r}} \right)$ (the global minimum point) may gives rise to two communities rather than the core and the periphery. But we cannot state that the network has no CP structure, it is possible that the CP structure is encoded in communities. To this end, we can check whether there are some local minimums in the curve of the motif conductance besides the global minimal point, and these local minimums can be used to detect whether the CP structure exists in the community. Therefore, we need to define the local minimum for the discrete sequence $x=\left\{ {{x_1},{x_2}, \cdots ,{x_n}} \right\}$: we call a point ${x_i} \left( {k < i \le n - k} \right)$ a local minimum of the function $h\left( x \right)$, if
 \begin{eqnarray}\label{9}
h\left( {{x_i}} \right) \le h\left( {{x_j}} \right)\;\;\left( {i - k \le j \le i + k} \right),
\end{eqnarray}
here we choose $k=3$.
%

For convenience, we define ${S_G}$ is a set whose elements are composed of the unhandled subgraph and ${S_C}$ is a set to save core nodes.  Algorithm 2 is developed by extending Algorithm 1.\\

\textbf{Algorithm 2:} (Input: graph $G$ and motif $M$. Output: core set ${S_C}$)

\textbf{Step 1:} Initialize  ${S_G} = \left\{ G \right \}$, ${S_C} = \emptyset$;

\textbf{Step 2:} If ${S_G}$ is an empty set, then: the algorithm ends and returns ${S_C}$, else removes an element (subgraph) $\tilde{G}$ from the set ${S_G}$, i.e., ${S_G} = {S_G}/\{\tilde{G}\}$.
$ \tilde{G}$ is the first element of the set ${S_G}$;

\textbf{Step 3:} Two sets are obtained by Step 1 - Step 5 of Algorithm 1 for subgraph ${\tilde{G}}$, that is $S$ and $\bar S$;

\textbf{Step 4:} If  all nodes' degrees of set $S$ are greater than the average degree, then ${S_C} = {S_C} \cup \left\{ S \right\}$, and go to Step 2;

\textbf{Step 5:}  If all nodes' degrees of set $\bar{S}$ are greater than the average degree, then ${S_C} = {S_C} \cup \left\{ {\bar S} \right\}$, and go to Step2;

\textbf{Step 6:} If there exists a local minimum in motif conductance function of sequence $S$ , then ${S_G} = {S_G} \cup \left\{ {{G_S}} \right\}$; If there exists a local minimum in motif conductance function of sequence $\bar S$, then ${S_G} = {S_G} \cup \left\{ {{G_{\bar S}}} \right\}$. Go to Step 2.

One point should be addressed: a network has no CP structure if all local minimums in the curve have been checked, namely, each dichotomization based on the local minimum cannot find CP structure.

We implement Algorithm 2 on two real networks. The first network is the Dolphin social network, which consists of 62 nodes representing the dolphins and 159 links denoting the frequent associations between dolphins~\cite{newman2004finding}. The network is composed of two communities. Fig.~\ref{fig6}(a) indicate that the two sets obtained from the smallest motif conductance ${\phi _M}\left( {{S_r}} \right)$  (first dichotomization) cannot give rise to CP structure, since neither of them can ensure that their degrees are larger than the average degree of the network. However there are local minimums on both sides of the smallest  motif conductance. We then respectively dichotomize the two local points to check whether there are cores in both sets (as shown in the inset of Fig.~\ref{fig6}(a)). Fortunately, two cores (red nodes in the inset of Fig.~\ref{fig6}(a)) are detected based on the second dichotomization regarding to the two local minimums. As illustrated in Fig.~\ref{fig6}(b): the red and purple nodes denote these two cores.

The second network is the Political blogs network (PB), which has 1222 nodes and 16714 connections in the network~\cite{ada2005pol}. The nodes of this network are blogs about US politics and the edges are hyperlinks between these blogs. This network displays a marked division into groups of conservative and liberal blogs, and has been viewed as a typical example of community structure~\cite{zha2006ide}. Fig.~\ref{fig6}(c) indicates that the first dichotomization based on the smallest motif conductance cannot find the core and the periphery. However, the second dichotomization based on the two local minimums can detect two cores (as shown in the inset of Fig.~\ref{fig6}(c)), corresponding to the red and purple nodes in Fig.~\ref{fig6}(d).

\subsection{Performance in synthetic networks} \label{sec4.3}

We have checked our algorithms in some real networks, however, the real partition of CP structure for these networks are unknown. As a result, the performance of different algorithms is hard to compare.  For this purpose, we consider the performance of these different algorithms in four classes of synthetic networks with one or two cores.

The synthetic networks are generated by using stochastic block models~\cite{zhang2015identification}. According to the definition of CP structure, the connection probability among core nodes ($P_{CC}$) should be larger than or equal to the connection probability between core nodes and peripheral nodes ($P_{CP}$), and further larger than  the connection probability among peripheral nodes ($P_{PP}$). Therefore, four synthetic networks with one or two cores are generated on the basis of $P_{CC}\geq P_{CP}>P_{PP}$ . The first synthetic network is constructed with one core, where the number of core nodes is $|C|=50$ and the number of the peripheral nodes is $|P|=150$. Moreover, we let $P_{CC}=P_{CP}=\theta$ and $P_{PP}=0.05$. The CP structure becomes more significant with the increase of the value of $\theta$. The second synthetic network is very similar to the first synthetic network, the only difference is that $P_{CP}=\frac{3}{5}\theta$ in the second synthetic network, since $P_{CC}>P_{CP}$ in many real networks CP structure~\cite{borgatti2000models,rombach2014core}. The third synthetic network is generated with two cores, where $|C_1|=|C_2|=50$ and $|P_1|=|P_2|=150$ are the size of two cores and the size of two peripheries. For each CP pair, we set $P_{CC}=P_{CP}=\theta$ and $P_{PP}=0.05$. Meanwhile, the inter- connection probability between two CP pairs is 0.01. The last type of synthetic network is slightly different from the third synthetic network: $P_{CP}=\frac{3}{5}\theta$.

Since the investigation of the detection of CP structure is still in its initial stage, the typical detection algorithm is few, moreover, many algorithms were proposed to detect single CP structure and the number of core nodes should be given in advance. Thus, few algorithms can be used to fairly compare.  Very recently one algorithm aimed at detecting multiple CP pairs was proposed by Sadamori Kojaku and Naoki Masuda (termed as KM algorithm). In this algorithm, authors design an algorithm by maximizing their defined quality function. Moreover, they remove some nodes as residual nodes by checking the statistical significance of each CP pair. Here we do not check the statistical significance of the CP structure when we implement this algorithm, since the statistical significance is not considered in other algorithms. In their work, they compare their algorithm with two other algorithms. One is BE-KL algorithm, which aims to detect a single core-periphery pair by maximizing $Q_{BE}$ (a quality function based on the Pearson correlation coefficient to measure the similarity between the given partition and its idealised CP structure ) using the Kernighan-Lin algorithm~\cite{kernighan1970efficient}.  BE-KL algorithm mainly focuses on how to detect single CP structure. The other algorithm is termed the two-step algorithm, the network is first divided into non-overlapping communities by maximizing modularity using the Louvain algorithm communities~\cite{blondel2008fast}, then the core and periphery in each community is detected by BE-KL algorithm again. Therefore, we compare our algorithm with the three algorithms: BE-KL, two-step and KM algorithms in the four types of synthetic networks.

We further introduce the normalized mutual information (NMI) to measure the performance of different algorithms, which is defined as~\cite{pizzuti2012multiobjective}:

 \begin{eqnarray}\label{9}
NMI(A, B)=\frac{2I(A,B)}{H(A)+H(B)}.
\end{eqnarray}
Here $A$ and $B$ are the partition determined by algorithms and the real partition, respectively, $I(A,B)$ is the mutual information of $A$ and $B$. $H(A)$ and $H(B)$ are the entropy of $A$ and $B$, respectively. NMI is in the range of $[0,1]$ and equals 1 only two partitions are total coincident.

\begin{figure}
\begin{center}
\includegraphics[width=3.2in]{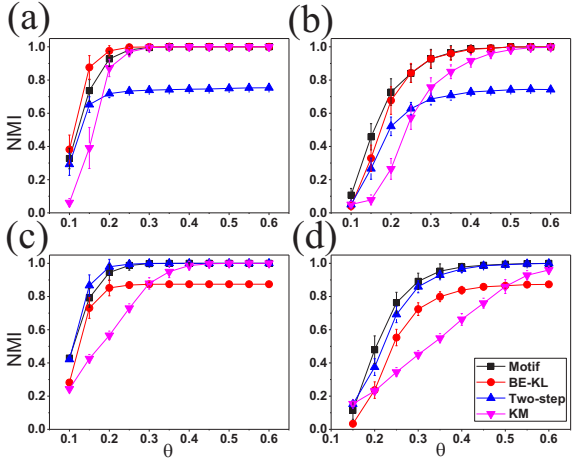}
\caption{The values of NMI as functions of $\theta$ for different algorithms are implemented in four types of synthetic networks. (a) $|C|=50$, $|P|=150$, $P_{CC}=P_{CP}=\theta$ and $P_{PP}=0.05$. (b) The parameters are all the same to (b) except for  $P_{CP}=\frac{3}{5}\theta$.  (c) $|C_1|=|C_2|=50$, $|P_1|=|P_2|=150$, $P_{CC}=P_{CP}=\theta$ and $P_{PP}=0.05$. Meanwhile, the inter- connection probability between two CP pairs is 0.01. (d) The parameters are all the same to (c) except for  $P_{CP}=\frac{3}{5}\theta$. The error bars
indicate the standard deviation.}
\label{synthetic}
\end{center}
\end{figure}

The comparisons of the four algorithms in four types of synthetic networks are illustrated in Fig.~\ref{synthetic}. Some phenomena can be observed: first, the performances of different algorithms become better when the value of $\theta$ is increased. The performance of KM algorithm is not efficient since the significance testing in the algorithm is discarded here, and moreover significance testing often leads to substantial time complexity. Second, the performances of BE-KL algorithm and our algorithm are most remarkable for synthetic networks with one core (see Figs.~\ref{synthetic}(a) and (b)). In detail, the performance of BE-KL algorithm is the best when $\theta$ is very small and $P_{CC}=P_{CP}$ , and the performance of our algorithm is very close to that of  BE-KL algorithm (see Fig.~\ref{synthetic}(a)). However, the performance of our algorithm is the best when  $P_{CC}>P_{CP}$ (see Fig.~\ref{synthetic}(a)). Third,  the performances of two-step algorithm and our algorithm are the most remarkable for synthetic networks with two CP pairs (see Figs.~\ref{synthetic}(c) and (d)). The performance of two-step algorithm is slightly better than our algorithm when $P_{CC}=P_{CP}$ (see Fig.~\ref{synthetic}(c)), and our algorithm yield better performance when $P_{CC}>P_{CP}$ (see Fig.~\ref{synthetic}(d)).
One should note that all three algorithms try to improve the similarity between the given partition with the idealised CP structure, this effect is often discounted as many real networks are far from the idealised CP structure. However, our algorithm detects the CP structure from the perspective of structure (i.e., motifs) rather than by maximizing the related quality functions.

\subsection{Detection of global CP structure by joining a leader node} \label{sec4.3}

\begin{figure}
\begin{center}
\includegraphics[width=3.2in]{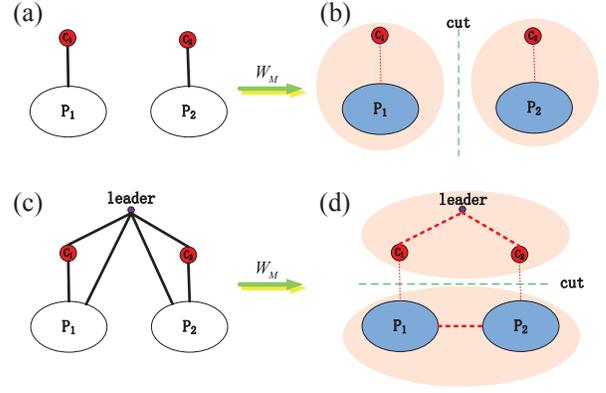}
\caption{Schematic illustration is given to demonstrate that the global core can be detected after joining one leader node into networks. (a) A network with two communities, where two cores $C_1$ and $C_2$ are encoded in two communities. $P1$ and $P2$ denote two peripheries. (b) the network is dichotomized two separated parts, where core nodes cannot be detected. (c) One leader node is joined into the network, which connects with all nodes in the network. (d) The global core and its periphery can be detected based on Algorithm 1 once one leader node is joined.}
\label{fig7}
\end{center}
\end{figure}

\begin{figure}
\begin{center}
\includegraphics[width=3.2in]{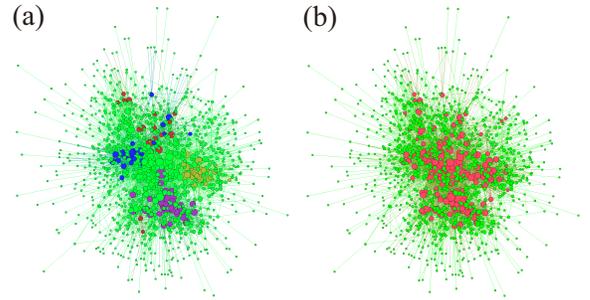}
\caption{Visualization of the detection of CP structure in email network. (a) Based on Algorithm 2, where several local cores are detected, marked by different colors. The green nodes are the peripheral nodes. (b) By adding a leader node, where a global core is detected (Red color). Green nodes are peripheral nodes. The sizes of nodes denote their degrees.}
\label{fig8}
\end{center}
\end{figure}

Take a schematic illustration in Fig.~\ref{fig7}(a) as an example, there are two cores $C_1$ and $C_2$ in two separated communities, dichotomization based on Algorithm 1 gives rise to two communities, but the cores in communities cannot be detected (see Fig.~\ref{fig7}(a)). Therefore, we further developed Algorithm 2 to detect multiple cores. For example, by implementing Algorithm 2 on the Email network~\cite{guimera2003self}, several local cores are detected in the network (Fig.~\ref{fig8}(a), green nodes are peripheral nodes and the nodes labeled by other colors are the different cores). However, the integrality of the network is destroyed due to the multiple dichotomizations (see Fig.~\ref{fig8}(a)). Sometimes, we are more concerned with whether there is a global core (or more precisely, ``hidden'' global core) from the whole network perspective. To do this, we need to design a method to detect the global core, no matter whether the network is a single CP structure or multiple CP structure structure network.

To overcome the difficulty in Algorithm 1, we can add a leader node into the network, which connects with all the nodes in the network (see Fig.~\ref{fig7}(c)). The motivation is that, once one leader is added into the network and with the topological transformation function $W_M$, the weight among core nodes and the weight among peripheral nodes are respectively increased, but the weight between core nodes and peripheral nodes is increased slightly. Now the dichotomization based on Algorithm 1 can detect the global core and its periphery (see the schematic illustration in Fig.~\ref{fig7}(d)). More importantly, such a global core still includes the local cores in different communities. As shown in Fig.~\ref{fig8}(b), the global core (red nodes) is obtained in the Email network after adding a leader node, which not only contains the local cores detected by Algorithm 2 but also contains some peripheral nodes who are misclassified due to multiple partition.

\section{Conclusions} \label{sec5}

\begin{figure}
\begin{center}
\includegraphics[width=3.2in]{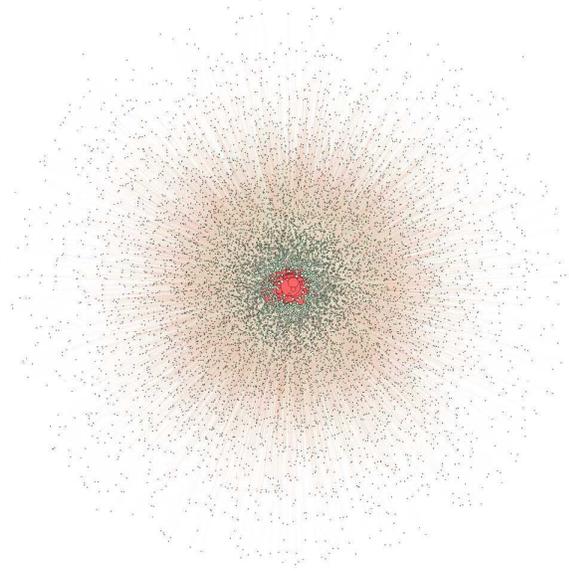}
\caption{Detection of CP structure in BlogCatalog network based on Algorithm 1, where red nodes and green nodes are core and periphery, respectively. The size of each node denote its degree.}
\label{fig9}
\end{center}
\end{figure}

In this paper, we have defined a 3-tuple motif to detect CP structure, and then the motif adjacency matrix based on the 3-tuple motif is constructed. Finally, the detection of CP structure can be realized by the smallest motif conductance, which is obtained by applying spectral clustering on the motif adjacency matrix. Our method has the following advantages: 1) our method can detect not only single or multiple CP structure, but also local or global CP structure; 2) our method is fast and can be applied large-scale networks. The complexity of our algorithm mainly depends on the computations of the motif adjacency matrix and an eigenvector. As stated in the supplementary material of Ref.~\cite{benson2016higher}, the motif adjacency matrix can be computed in $O\left( {{{\bar d}^3}n} \right)$ time , and the eigenvector can be computed in $O\left( {(m + n){{(logn)}^{O\left( 1 \right)}}} \right)$ time by using fast Laplacian solvers~\cite{trevisan2013lecture}. Our algorithm is fast, for instance, the CP structure in the BlogCatalog network~\cite{tang2009relational} can be quickly detected by our method (see Fig.~\ref{fig9}).  This network has 10312 nodes, 333983 links and the average degree is 64.78. It takes 1856.85 seconds when using MATLB2015b to implement our algorithm on a PC with Inter Core i7-4790 CPU 3.60GHZ and the Windows 7 64bit operating system; 3) our algorithm is parameter free.

\ifCLASSOPTIONcaptionsoff
  \newpage
\fi

\end{document}